\documentclass[twocolumn]{aastex631}

\usepackage{tikz}
\usetikzlibrary{decorations.pathreplacing}
\usepackage{amsmath}
\usepackage{amssymb}
\usepackage{amsfonts}
\usepackage{textcomp}
\usepackage{gensymb}
\usepackage{multirow}
\usepackage{float}
\usepackage[hang,flushmargin]{footmisc}
\usepackage{microtype}
\usepackage{graphicx}
\usepackage{tabularx}

\shorttitle{Simulated impact on LSST}
\shortauthors{Kandula, Tyson, Askari, Fankhauser}
\begin{document}

\title{Simulated impact on LSST data of Starlink V1.5 and V2 satellites}


\author[0009-0009-8681-5318]{Phanindra Kandula}
\affiliation{Department of Physics and Astronomy, University of California \\
One Shields Avenue \\
Davis, CA 95616, USA}
\affiliation{currently at University of Massachusets, Amherst, MA}

\author[0000-0002-9242-8797]{J. Anthony Tyson}
\affiliation{Department of Physics and Astronomy, University of California \\
One Shields Avenue \\
Davis, CA 95616, USA}

\author[0009-0008-3846-9708]{Jacob Askari}
\author[0009-0008-2880-4752]{Forrest Fankhauser}
\affiliation{SpaceX \\
One Rocket Rd. \\
Hawthorne, CA 90250, USA
}

\begin{abstract}

In order to reduce the impact on ground-based optical astronomy, the new Starlink V2 satellites incorporate improvements to the chassis brightness through dielectric mirrors, off-pointing solar arrays, and black paint on exposed components.  To assess the effectiveness of these mitigations for the general case in which the reflectivities are initially unknown, we simulate LSST operations and repeated photometry of every satellite in simulated model constellations. We derive a brightness model of the Starlink V2 satellite and study the simulated apparent brightness as a function of the satellite position relative to the observer and the sun. We find that the V2 Starlink satellites appear brightest at two distinct positions in the sky: when oriented toward the sun at low elevations where light is specularly reflected, and nearly overhead where the satellite is closest to the observer.

A simulation of Starlink V2 satellites at 550 km height distributed across a series of Walker constellations~\citep{walker1984satellite}, 
with varying inclinations
was analyzed to study the impact on the LSST observations. The results of the V2 Starlink satellite trail simulation were compared to a similar simulation of V1.5 Starlink satellites.

We find that some bright satellites will be visible in LSST observations. For every thousand V1.5 Starlink satellites imaged by LSST in the first hour of a summer night, we find 1.2 of them will appear brighter than 7 AB magnitude. By comparison, for every thousand V2 Starlink satellites observed, we find only 0.93 of them will appear this bright.
The off-pointed solar array and reduced diffuse reflection of the chassis mitigate the brightness.

 Finally, we simulate lowering this Walker constellation to 350km. Only 0.56 V2 Starlink satellites per thousand brighter than 7 AB magnitude will be observed in the first hour at this height. This is a $\sim$40\% reduction in number of bright satellites entering the focal plane compared to the constellation at 550km height. 
 We find that a combination of factors yield an apparent surface brightness of these satellites for LSST operations only 5\% brighter than at 550km orbit.

\end{abstract}


\newpage

\section{Introduction}

Satellite constellations in low Earth orbit (LEO) pose a challenge to ground-based astronomy, particularly wide field surveys such as the NSF-DOE Vera C. Rubin Legacy Survey of Space and Time (LSST)~\citep{ivezic2019lsst,tyson2020mitigation, hu2022satellite}. 
This is due to the rapid and wide sky coverage of LSST. Each exposure is expected to have at least one LEO satellite trail in the 10 sq.degree focal plane, with many more near twilight and low in the sky.  In a previous study~\citep{fankhauser2023satellite} a satellite brightness simulation package LumosSat~\citep{lumos2023}  was developed and used to predict satellite brightness in two cases: with and without prior knowledge of the Bidirectional Reflectance Distribution Function (BRDF) of the satellite components.  It was found that with sufficient re-observation of each satellite it is possible to derive a best fit BRDF model which can then be used to predict satellite brightness given the sky location relative to the sun.  

In this paper we develop this technique further, comparing two models of Starlink V2  satellites 
with Starlink V1.5 satellites. These simulations are calibrated using repeated ground-based photometry of each V1.5 satellite.  We use these data to derive the BRDF model, and then compare the predicted brightness distribution with the measured value.  The model is optimized and then used in a simulation of thousands of V2 satellites in two simplified orbital configurations at different altitudes. It is difficult to predict the total quantity of Starlink satellites on-orbit during LSST operations, so all metrics are reported as per-thousand on-orbit satellites. This allows our results to be easily scaled to different constellation sizes. Using the LSST observation scheduler,
~\citep{2019AJscheduler,scheduler2025}, 
the resulting model is used to predict V2 satellite tracked magnitude vs time and solar angle during the night.

\section{Photometric monitoring data}
\texttt{Slingshot Aerospace} tracked v2 satellites and recorded their brightness along the orbit. Satellite position and time of observation were also provided and corroborated with SpaceX telemetry. Measurements were made using remote scheduling from multiple observatories in June 2023. SpaceX parsed down measurements to only include nominal, on-station conops. These observations utilized horizon to horizon tracking. The Slingshot telescope gimbals to track the satellite as it comes above the horizon, passes overhead, then passes under the horizon. The precision of the Slingshot photometry is very high, much higher than the observed variations in flux of the satellites. These data are available in our open data repository referenced below. Data used for v1.5 was measured with different methods and observatories per a previous study~\citep{fankhauser2023satellite}.

\section{Effect of satellite altitude}
It is expected that the orbit altitude of a LEO satellite constellation will impact ground based observations.  While the satellite will appear brighter at lower ranges (a $1/r^2$ effect), the image of the satellite moves across the LSST focal plane faster (proportional to $r$). The net effect is that the surface brightness of the trail increases as $1/r$.  The surface brightness is important, because there is a surface brightness threshold for trail detection.  

The length of the trails also increase by the same factor.   When combined with the LSST observing schedule on the sky, it is expected that lower altitude constellations may have less impact on LSST data because they may evade detection. Most of the brighter satellites would be detectable only below the elevations used in the scheduled LSST observations.

In this paper we therefore simulate two versions of V2: V2a at 550 km and V2b at 350 km altitude.  As described below, both simulations use a simplified Walker constellation~\citep{walker1984satellite} at a 45 degree inclination. 

\bigskip

\section{Satellite Model}
    The Starlink V2 satellite model we developed for our simulation is a series of surfaces that make up the satellite. Each surface of the satellite uses different BRDF models to calculate the intensity of light reflected based on the incoming sun ray and reflected angle. The total brightness of the satellite is given by the sum of each surface’s reflected light in the direction of the observer. By using the BRDF specifications of each material, a reflection model was fit to each surface. To compensate for the brightness from other small surfaces and brightness from other sources, a free parameter in the form of a surface using the Phong reflection model~\cite{phong1973} was added to the chassis. The parameters for the Phong model are adjusted so that the final satellite brightness is closest to the measured brightness data obtained from the monitoring service \texttt{Slingshot Aerospace}. A custom inverse-log error function is used while fitting all surfaces to ensure that brighter satellites are weighted more. The bus mounted black components (including reaction wheels and pantograph, which is the solar array deploy mechanism) and the chassis contribute most to the total brightness of the satellite as seen by the observer. For simplicity, all bus mounted black components are grouped together in the brightness model into a single surface. This surface is referred to as the pantograph, which is one of the largest bus mounted black components but contributions from other bus mounted black components are grouped into this surface. 
    
\begin{figure}[h]
    \centering
    \includegraphics[width=0.46\textwidth]{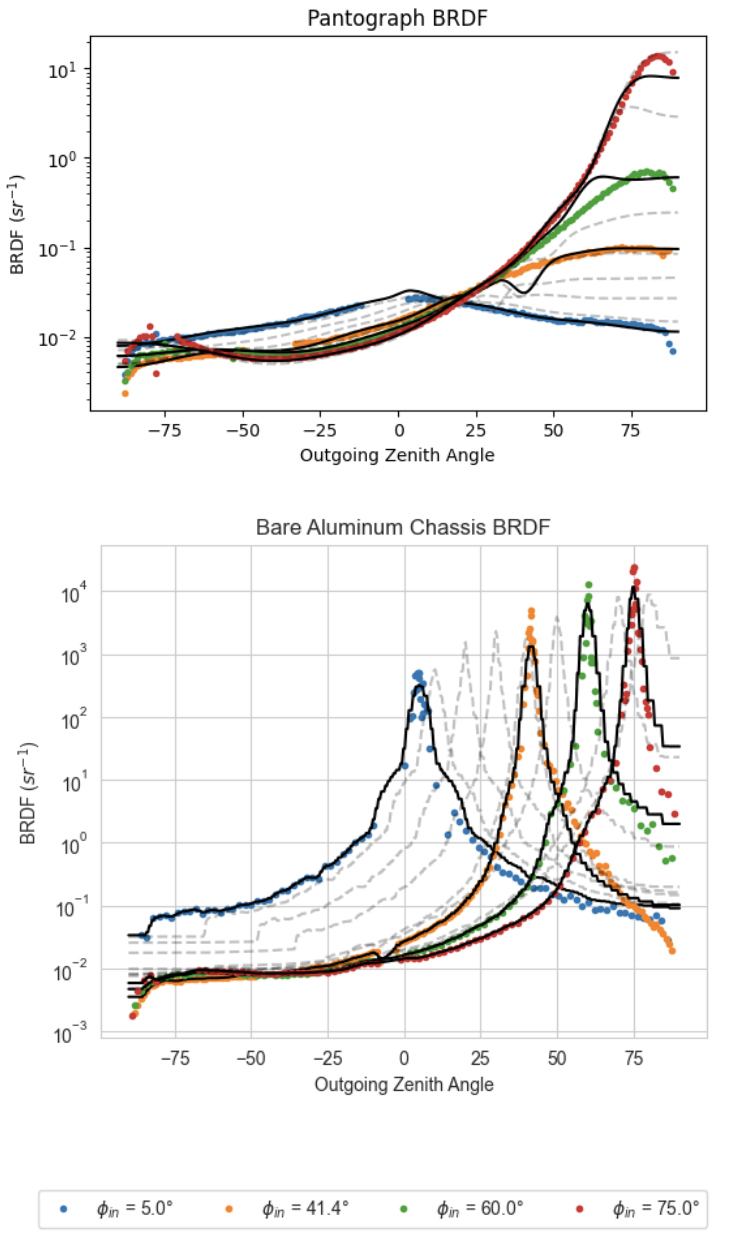}
    \caption{Measured BRDF data and interpolated BRDF of the aluminum in the surface of the chassis and pantograph paint in 1D. The measured BRDF is presented as a scatter plot for the four specific incoming zenith angles. While the interpolated BRDF model is depicted as a continuous function shown by the black line. The gray dashed lines represent the interpolated BRDF model for incoming zenith angles incremented by 10 degrees. The angle of the specular reflection equals the angle of incidence for the aluminum. This does not apply to the black paint used for the pantograph, making it harder to model accurately.}
    \label{ChassisAndPantographBRDF}
\end{figure}

    BRDFs of these surfaces along with their prediction models are plotted in Figure~\ref{ChassisAndPantographBRDF}. Each surface follows the law of reflection where the angle of incidence and the angle of reflection are the same. The reflected angle is measured on the opposite side of the surface normal, creating a mirror-like symmetry between the incident and reflected rays. However, the pantograph (and other bus mounted components) are coated with a unique black paint that does not follow this principle. This is seen in the top graph of Figure~\ref{ChassisAndPantographBRDF}.  While it is not perfectly represented by the binomial reflection model, the model still encapsulates the general pantograph BRDF behavior.  Only when the outgoing zenith angle matches the incoming zenith angle, does the model predict inaccurately. When the incident zenith angle is near -41.4°, the model exhibits the greatest error at the matching outgoing angle of reflection. Fortunately, few satellites that enter the area of the sky where the model's incoming zenith angle matches the outgoing zenith angle.  On the other hand, the model for the chassis' bare aluminum takes a more meticulous approach by using a inverse-distance interpolation method on the lab data to get a brightness prediction. The chassis surface primarily consists a dielectric mirror, assumed to cover 97.5\% area, while the remaining 2.5\% of the area is assumed to be bare exposed aluminum. There are some areas where dielectric mirror application is not feasible and assuming some area has the more diffuse optical properties of bare aluminum helps account for that and any imperfections in the dielectric mirror.  Most light that hits the dielectric mirror is reflected back to space, making the exposed aluminum surface the biggest contributor to the chassis' overall brightness as seen by the observer. To prevent over fitting and the model slowing down, the interpolation method was not used for the pantograph. 

\subsection{Model Performance}
\begin{figure}[h]
    \centering
    \includegraphics[width=0.46\textwidth]{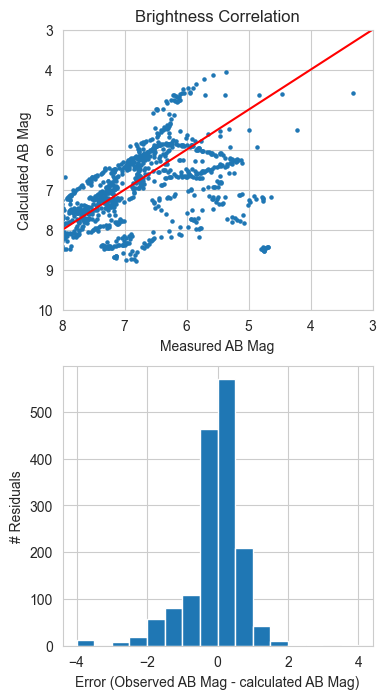}
    \caption{Top: V2 Starlink Satellite predicted brightness vs measured brightness. The scatter increases for brighter satellites; but the majority of the satellites remain near the red line representing a perfect fit. Bottom: The prediction error is defined as Measured AB minus Calculated AB magnitude. The model's least accurate predictions are underestimating the brightness of the satellite.} 
    \label{Satelliteresidual}
\end{figure}

The model is validated against the V2 Starlink observations by \texttt{Slingshot Aerospace}.  Using this as data to fit, we then computed the theoretical brightness of these satellites using the Lumos software. Figure \ref{Satelliteresidual} shows the brightness correlation between the predictions and the observed brightness of the satellites captured by \texttt{Slingshot Aerospace}. The model’s predictions are roughly accurate. The rms error between predictions and observed brightness is 0.74 AB magnitude. As satellites get brighter, the model’s accuracy falls, making the brighter satellites contribute most of the error.

\begin{figure*}

    \centering
    \includegraphics[width=1\textwidth]{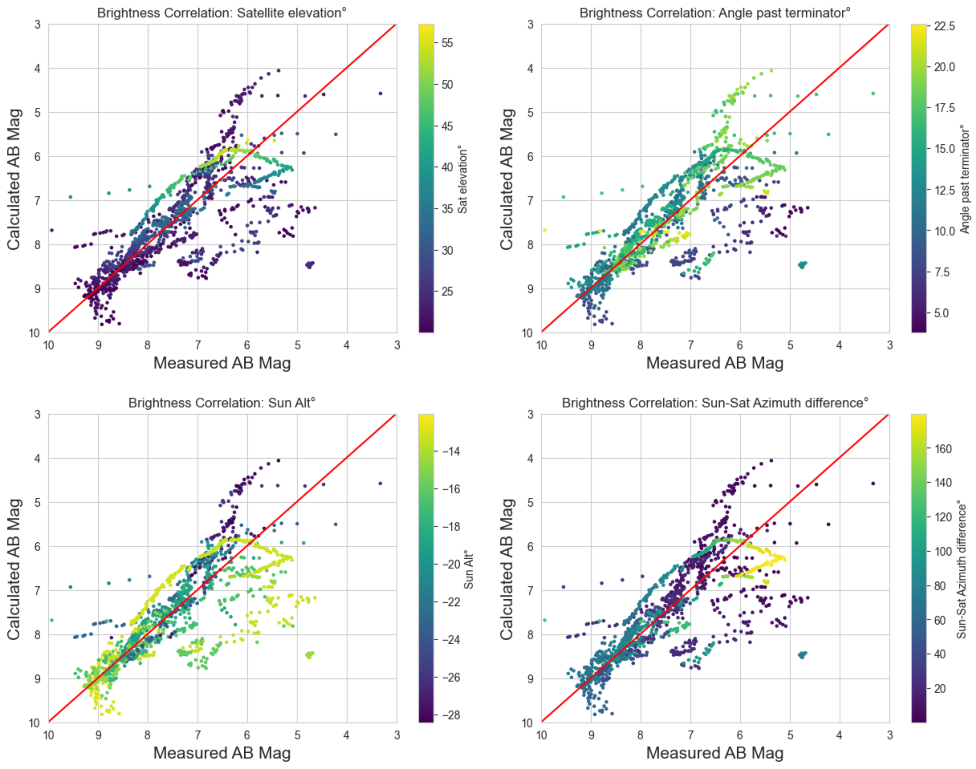}
    \caption{Brightness correlation between the satellite’s predicted brightness and the Sun-Satellite azimuth difference, satellite altitude, sun altitude and satellite’s angle past the terminator. Given that the model’s prediction error is larger for satellites observed to be brighter than 6AB, the sun-satellite-observer configurations for the brightest satellites are where the model struggles to predict brightness. In all cases, the satellites with the largest errors have two or more factors that increase the brightness, most notably the satellites that are towards the sun and at low elevations.}
    \label{brightnesscorrelation}
\end{figure*}

Studying the correlation between the satellite’s position and the satellite’s error in Figure \ref{brightnesscorrelation} reveals that the model’s accuracy is worse at the areas of the sky prone to bright satellites. We consider the dependence on satellite azimuth as seen by the observer. Sun-Sat azimuth difference is defined as the angular difference between the sun’s azimuth and the satellite’s azimuth from the observer’s perspective. Satellites with a low Sun-Sat Azimuth difference that are observed by \texttt{Slingshot Aerospace} are very bright. The model struggles to predict the brightness of satellites that are within 25 degrees of the sun’s azimuth. These satellites are almost in the same plane as the observer and the sun, making it easier to visualize these satellites in 2D as shown below in Figure \ref{notspecular}. The rms error for these satellites is 0.89 AB mag, while the rms error for satellites with larger Sun-Sat azimuth difference is 0.63 AB magnitude. Bright satellites in the direction of the sun (low Sun-Sat azimuth difference) suggest forward scatter off the chassis.
   
\begin{figure}
    \centering
    \includegraphics[width=1.0\linewidth]{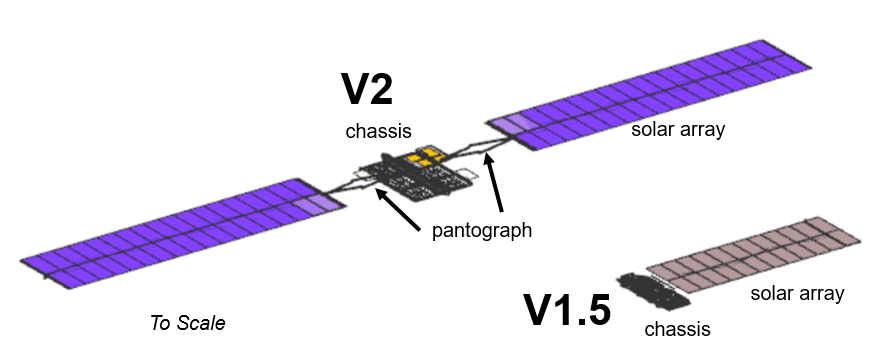}
    \caption{Starlink satellite configurations covered in this paper. The chassis refers to the main satellite body that remains pointing nadir. The solar array can articulate. On v1.5, the solar array articulates to track the sun. On v2, the solar array tracks the sun, except in brightness region of concern, where it off-points to mitigate brightness. The pantograph, which deploys the solar array, is mostly behind the solar arrays with some length exposed between the chassis and the solar array on v2.}
    \label{fig:Starlink v1.5 and v2 Satellites}
\end{figure}
The V1.5 and V2 satellites are compared to scale in Figure~\ref{fig:Starlink v1.5 and v2 Satellites}. The V2 Starlink solar array is articulated to mitigate solar array brightness, while also optimizing for many other parameters. The solar array is offpointed such that observers on Earth have minimal line of sight to the illuminated front of the solar array. This results in roughly a 10$\%$ reduction in satellite power generation for the currently implemented concept of operations. The Starlink v2 solar arrays were sized appropriately to tolerate this reduced power generation. For simplicity of modelling, it is assumed that the solar array is angled 45 degrees to the chassis and away from the sun. As much of the pantograph as possible is kept hidden behind the solar array; however, the remaining exposed pantograph is angled to the chassis but towards the sun so that light is scattered to earth. The pantograh is painted matte black so most light is absorbed and the remaining scatter is diffuse.  Other bus mounted components that are angled similarly are also painted matte black. The pantograph and other bus mounted components contribute significant brightness for satellites at higher elevations while the chassis is the largest contributor of brightness for satellites at low elevations. 

Similarly, satellite elevation plays an important role in the apparent brightness of the satellite and the prediction’s accuracy. Satellites with large prediction errors are typically observed at elevations below 30 degrees. The model’s prediction accuracy improves as the satellite goes higher in the sky. The fainter low-elevation satellites that are dimmer than 7 AB magnitude are predicted more accurately by the model. The rms error for bright low-elevation satellites is 0.97 AB magnitude while the rms for the fainter low-elevation satellites is 0.79 AB magnitude. These bright low-elevation satellites are all found in the direction of the sun. The satellite’s brightness and the model’s prediction accuracy are both dependent on the satellites position relative to the observer.
For all the surfaces used in the satellite, the specularly reflected light ray is always in the same plane as the incoming light ray relative to the surface. This explains why the satellites that are in the direction of the sun are usually much brighter than satellites at other angles. However, these satellites do not specularly reflect to the observer as visualized in Figure \ref{notspecular}. The outgoing light ray represents the specularly reflected light ray. Instead, the ray is diffused at nearby angles which do reach the observer. So although the angles are misaligned for a specular reflection, these problematic satellites are still bright enough to be mistaken for a specularly reflected satellite. Furthermore, if there are other small unmodeled surfaces on the chassis, it could also contribute to the discrepancy between predictions and observed brightness. Specifically, there are parabolic antenna dishes that are painted black and gimbal to communicate with ground stations. Brightness contributions from these vary with gimbal position and are not represented in the satellite brightness model.

\begin{figure*}
    \centering
    \includegraphics[width=1\textwidth]{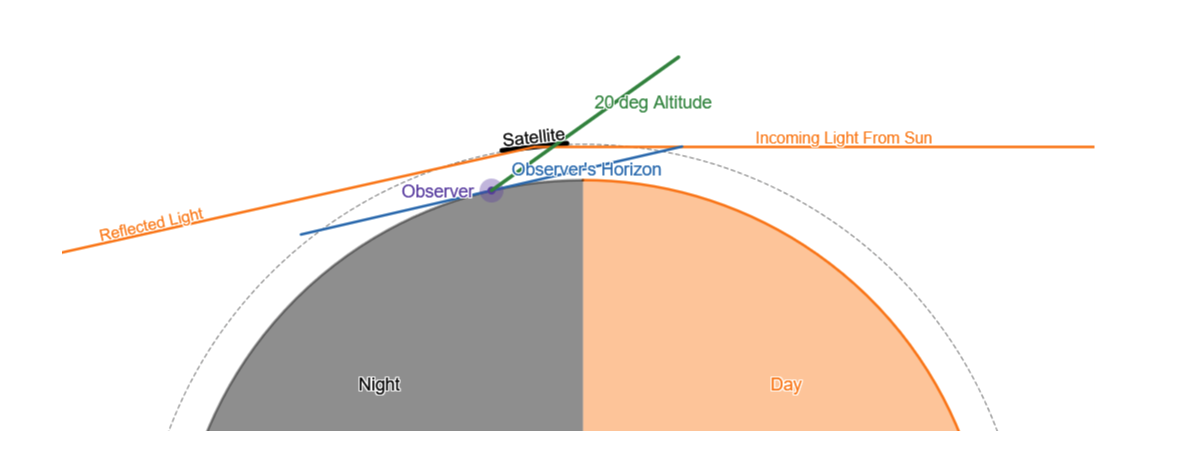}
    \vspace{-0.5cm}
    \caption{Bright Satellite Position. This brightness, caused by the sun being in the same direction as the satellite, creates the appearance of a specular reflection. Since the reflected light ray is directed into space, observers on Earth can only see the diffused light rays. By being at the same direction as the sun, the observer is closest to the specularly reflected light ray.}
    \label{notspecular}
\end{figure*}

\vspace{0.5cm}

\begin{table*}[ht]
\hspace*{-2 cm}
\centering
\resizebox{1.1\textwidth}{!}{
\begin{tabular}{|l|l|l|l|l|l|l|}
\hline
\textbf{Satellite Elevation°} & \textbf{Azimuth Difference°} & \textbf{AOI°} & \textbf{Sun Altitude°} & \textbf{Observed Brightness} & \textbf{Calculated Brightness} & \textbf{Error} \\ \hline
Low & Low & High  & Low  & Bright & Very Bright & Large \\ \hline
Low & Low & Low &  High  & Bright & Not as Bright & Large \\ \hline
High & High & High &  High  & Bright & Bright & Small \\ \hline
Low & High & - & - & \textgreater{} 7 AB & \textgreater{} 7 AB & Small \\ \hline
\end{tabular}
}
\caption{A summary of the model's effectiveness in a table format.  The accuracy of a model prediction for a satellite is listed. }
\label{satellite_brightness_summary}
\end{table*}

Satellites in Earth's shadow are not illuminated by direct sunlight.  A satellite’s brightness is dependent on the angle past the terminator. Since the chassis of the satellite is parallel to the earth, some quick geometry will show that the angle past the terminator and the angle of incidence (AOI) are the same for the chassis. In Figure \ref{brightnesscorrelation}, the model under-predicts the brightness when the satellite has a low angle past of terminator. Conversely it predicts more accurately when the satellite has an angle past terminator greater than 10°. Furthermore, the Figure \ref{brightnesscorrelation} shows that the model predicts poorly at low altitudes and small angles past the terminator. We also know the sun-sat azimuth difference also affects the predictions at small angles. So more specifically, the model struggles with under predicting the brightness when the satellite has a small angle past terminator, low elevation and a small sun-sat azimuth difference. 

Meanwhile the correlation of the sun’s altitude and the satellite’s brightness shows that the simulation over-predicts the brightness of the satellite when the observer’s sun sets to a low altitude of less than 24° in Figure \ref{brightnesscorrelation}. However these satellites do not have a low angle past the terminator. In fact, the observer has set far enough into the night that it is impossible to see a satellite at that low angle past the terminator. In summary, the model is poor at predicting the brightness of a satellite that has a low sun-sat azimuth difference, low elevation and either a low angle past terminator or a low sun altitude. Taking this into account when simulating these satellites, we can estimate how accurate the model’s predictions are for our three summer surveys. Table \ref{satellite_brightness_summary} contains a summary of how a satellite's position determines how accurate the prediction is.

 Double scatter from earthshine makes satellites brighter at lower elevations. Our model calculates earthshine for low-elevation satellites. Earthshine is calculated by dividing the surface of the Earth into 151 x 151 panels. The simulation calculates the light emitting from Earth panels that illuminate the satellite to calculate the BRDF from earthshine. The brightness of these panels are calculated from the previous surveys of brightness of the earth as seen by satellites. Calculating the brightness of these panels increases the computational time by ${O(n^2}$). Due to the computational cost of calculating earthshine, we only use it for the small number of satellites that are at a low elevation and not under the Earth's shadow.

\subsection{Brightness Distribution}

To further understand the model’s predictions, we simulated a sky full of V2 Starlink satellites and calculated the brightness of each satellite at various sun altitudes. There are satellites that are visible until the sun reaches an altitude slightly more than -45°. However, we only consider the brightness of satellites using sun altitudes between -12° and -33°. The earliest LSST observes is at nautical dusk and the latest is at nautical dawn. Since the LSST does not pursue observations below 20° elevation, all satellites that are in the observation area cease being visible at a sun altitude of -33°.

\begin{figure*}
    \centering
    \includegraphics[width=1\textwidth]{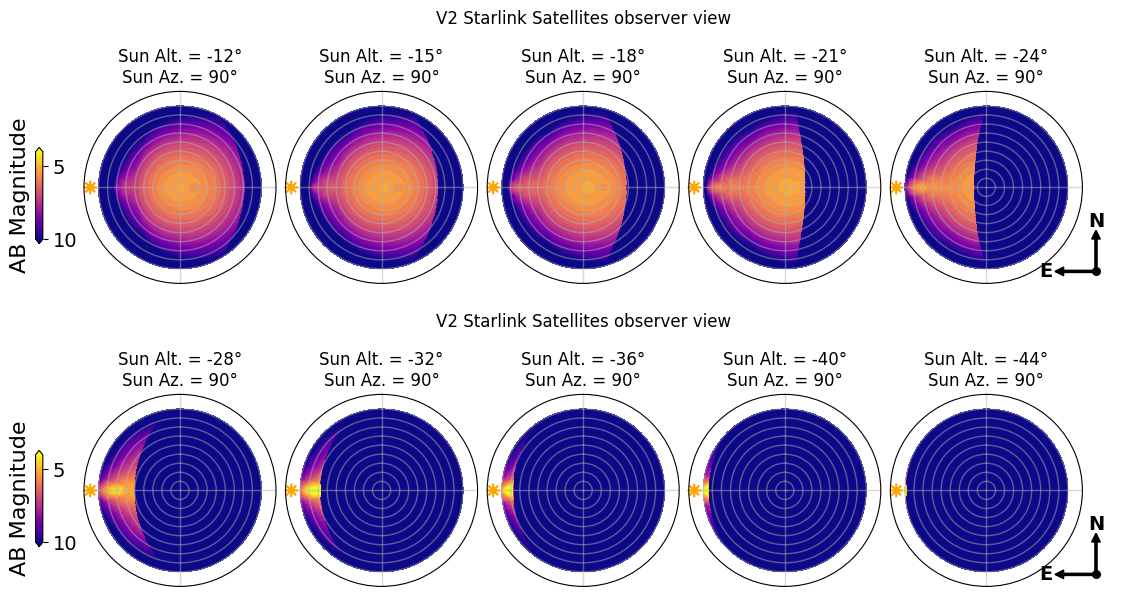}
    \caption{Brightness distribution of all satellites in morning twilight. The satellite's altitude (Alt) is its elevation above the horizon; its angle along a radius from the outer edge of the circle. The satellite's azimuth is represented by the polar angle. The origin on these plots are the satellites that are directly above the observer while the edges of the plot are the satellites at the horizon. There are polar grid lines at every 10° elevation and there are cardinal grid lines at 90°. 
    The sun is in the east (orange star) and represents morning twilight. The plot shows brightness distribution of satellites with sun altitude ranging from -12° to -44°. As the sun sets, more satellites fall within the shadow of the earth and more satellites start to specularly reflect. }
    \label{V2osberverview}
\end{figure*}

Diffusely reflecting components of satellites are typically brightest when they are nearly overhead and at their closest distance to the observer, since light intensity is inversely dependent on the distance between the observer and the satellite. For the V2 Starlink satellite specifically, the pantograph is responsible for the most of the brightness at higher elevations, as seen in Figure \ref{V2osberverview} . The pantograph is angled at 45° to the chassis but in the direction of the sun, causing all the reflected light to be directed towards the Earth. Due to the complexity of the BRDF of the black paint used for the pantograph, the brightness model is not a perfect fit.  While it captures the overall behavior of the pantograph, the small dip in brightness at high elevations are due to this. We avoided further enhancements to the brightness model to prevent overfitting.  

 Another condition under which a satellite is very bright is when the incoming sunlight is reflected specularly to the observer. This can happen only when the sun's altitude is near -45° and the satellite is at the horizon. Neither of these scenarios are within the LSST’s observing schedule. However, we simulate this arrangement to verify  this as shown in Figure \ref{V2osberverview}. As expected, the satellites that are directly overhead are the brightest whenever they are illuminated by the sun.

 As the sun's altitude continues to decrease in the top half of Figure \ref{V2osberverview}, more satellites go into the shadow of the earth. A new brightness peak begins to appear at lower satellite altitudes. This is the same subset of satellites which caused the model to over-predict brightness compared with what we observed: they are characterized by low satellite elevation above the horizon, in the direction as the sun, with the observer’s sun altitude low.  When we continue to go down in sun altitude past nautical twilight, we see that this brightness peak is the same specular peak that occurs when the sun altitude is at -44° before the satellites go into the earth’s shadow. This confirms that what the model calculates is a specular reflection. However, the specularly reflected light ray never actually reaches the surface until the sun altitude is extremely low. So the brightness peak we observe is a result of spill-over light from the specularly reflected light ray as it gets closer to the surface as the sun's altitude and satellite elevation gets lower.  We do know that the chassis of the {V}2 Starlink has a lower diffuse reflection, i.e, more light is confined within the specularly reflected light ray making satellites appear dimmer than the Gen 1.5 Starlink. This is discussed in detail below. Unfortunately, our model over-estimates the brightness in this scenario. 

\subsection{Model discrepancies}
\begin{figure}
   \centering
   \includegraphics[width=1.0\linewidth]{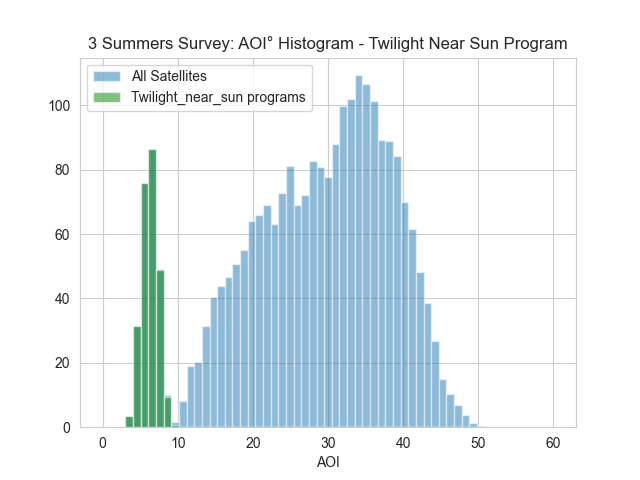}
   \caption{A histogram of the satellites’ angle of incidence distribution and distribution of the satellites observed through the twilight near-sun programs. There are only four programs that point in the direction of low angle of incidence satellites. They are all the twilight near-sun programs with the goal of finding potentially hazardous asteroids. 
   The y-axis is sum over three summers in units of 1000 satellites.
   }
   \label{twilightprograms}
\end{figure}

The LSST does not observe the night sky uniformly in elevation. Each program on the LSST scheduler serves a different purpose, with some emphasizing low elevations and some high elevations in a different part of the sky~\citep{2019AJscheduler,bianco2021optimization}. Since the model cannot accurately predict the brightness of satellites in some areas of the night sky, the impact of the poor predictions are localized to only those LSST programs that look at those areas. We know that the model under-predicts the brightness of satellites that have an elevation of less than 30°, are in the general direction of the sun such that the azimuth difference is less than 45°, and maintain an angle of incidence less than 10° as detailed in the previous sections. The programs that focus their observations in this area of the sky are more likely to encounter satellites whose brightness will be under-predicted. Interestingly, as their name suggests, the “twilight near sun” programs are the only programs that observe at very low altitudes and in the direction of low angle of incidence satellites. The “twilight near sun” programs are run during both morning and evening twilight. However they rarely observe below 30 deg elevation. Figure \ref{twilightprograms} shows the density of the satellites that enter the four “twilight near sun” programs’ field of view with respect to the satellite’s angle of incidence.  Other programs do not encounter these bright satellites, and the model’s predictions in those other programs are more precise. 
The “twilight near sun” programs are designed to look for potential hazardous asteroids (PHAs). Our model does not cover near horizon situations, resulting in  poor accuracy in detecting bright satellites at these small elevation angles; depending of PHA detection algorithms it is possible that satellites will increase the noise in the PHA detection. 

The model over-predicts brightness when the satellite has a low altitude, a sun-sat azimuth difference less than 20° and high angle of incidence. Only a few programs observe at low altitudes when not at twilight making the number of over-predicted bright satellites small. The “DD:XMM$\_$LSS” and “DD:ELISES1” are the most common ``LSST Deep Drilling'' programs that look into this area of the sky. However, unlike the “twilight near sky” programs, it is not intended to look this low as only 2$\%$ of their observations would have a satellite with over-predicted brightness.  
While it is important to remember that the model’s error of these satellites are sometimes more than 1.5 AB magnitude, these events are rare and will not affect the study’s findings given the large sample size of three summer's worth of satellite simulation data. We discuss this topic more in the results section.


\section{Results}
\begin{figure*}[ht!]
   \centering
   \includegraphics[width=1\textwidth]{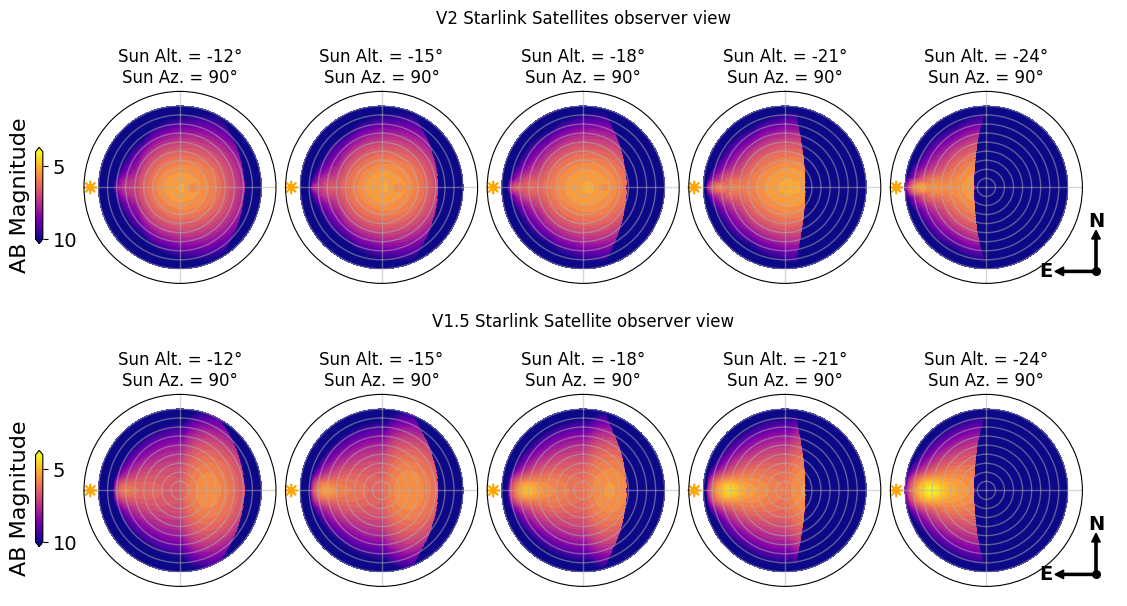}
   \caption{Brightness distribution comparison between the V1.5 and V2 models. Top: The brightness distribution for a V2 Starlink model. Bottom: The brightness distribution of the V1.5 Starlink model. Two distinct brightness peaks are visible. The first brightness peak at low elevation (near the outer edge of the sky circle)  is from specular reflection. The second brightness peak is at the opposite side of the sun caused by the back scattered light from the solar array. This does not happen in the V2 model.}
   \label{modelcomparison}
\end{figure*}

The {V}2 Starlink satellites' two main brightness mitigation methods that distinguish itself from Gen 1.5 Starlink satellites are the off-pointing solar array and reduced diffuse scatter from the chassis. The off-pointing solar array is angled to the chassis such that the incoming sun light is reflected back into space away from the observers. Due to these conops, which reduce power generation, the {V}2 Starlink satellites' solar array is invisible to observers on Earth. However, the Gen 1.5 Starlink satellites' have solar arrays perpendicular to the chassis reflecting sunlight to the observers through back scatter when the satellite is in the opposite direction of the sun. This back scatter is very apparent in the brightness distribution of the Gen 1.5 Starlink satellites as seen in Figure \ref{modelcomparison} for satellites that are in the opposite quadrant of the sun. 


Another area of the sky where both versions of the satellite are susceptible to appearing intensely bright is at low satellite elevation and in the direction of the sun. As mentioned before, the specularly reflected light ray causes this brightness peak. But with the chassis being less diffusive, this brightness peak is not evident until the sun sets to -21°. When the sun sets to -24°, the specular reflection can be seen with both satellite models, but the V1.5 Starlink satellite’s peak is brighter and covers a larger area of the sky. We know that our simulation model of the V2 Starlink satellite exaggerates this brightness peak, so the area of the sky that the V2 Starlink satellites reflect specularly is even dimmer and smaller than what is plotted. Other bright V2 Starlink satellites are expected to be seen at higher altitudes due to the satellite being closer to the observer rather than as a result of the specular reflection extending to the observer. We can also see that the V1.5 model’s brightness is more dependent on the azimuth difference between the sun and satellite because the two brightness peaks only occur when the azimuth difference between the sun and satellite is small. Since the brightness peak of the V2 Starlink satellite is nearly overhead and less dependent on the satellite’s azimuth, the overall brightness is also less dependent on the satellite azimuth.

We model a Walker constellation to be consistent with regulatory filings  
and simulate realistic satellites moving across the LSST focal plane. With a working V2 Starlink satellite model, we can simulate the survey and compare with the V1.5 model results.  The LSST scheduler contains each planned observation’s telescope pointing direction in RA-DEC coordinates, the time the observation starts and the associated program and filter. This simulation focuses on the g-r-i-z-y filters and ignores the u-filter. Since an observation lasts for at least 15 seconds, we calculate the position at 5-second intervals of the observation to interpolate the path of the trail. The trails’ brightness is calculated as the average of the satellite’s tracked magnitude in the focal plane. Brightness is calculated using the V2 Starlink model outlined before, and a Gen 1.5 Starlink model from another study \cite{fankhauser2023satellite}. 

\begin{figure*}[ht!]
   \centering
   \includegraphics[width=1\textwidth]{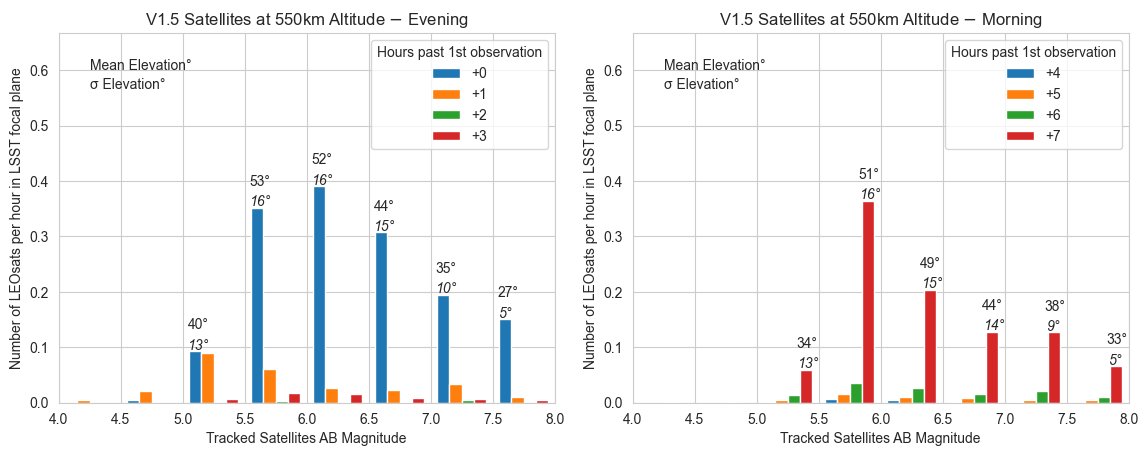}
   \caption{Density (per-thousand on-orbit satellites) of expected bright V1.5 satellite by the hour since/before the start of the night past/before nautical twilight. The evening twilight hours are on the left and the morning twilight hours are on the right. Most bright satellites are seen within the first hour of observation and the final hours of the observation in morning twilight. The mean and standard deviation of the satellite elevation for each brightness category are shown above each bar.  The mean brightness of all satellites within the first hour is 7.1 AB Magnitude with {many} satellites being between 5.5 - 6 AB magnitude. Within the first hour since the start of the observation, we expect to see 1.2 satellites per thousand brighter than 7 AB magnitude. On average, 0.1 per thousand of these bright satellites will be brighter than 5.5 AB magnitude. The fainter satellites are seen with a low mean elevation with minimal brightness variance, whereas the brighter satellites are at significantly higher mean elevations but exhibit greater variance in apparent brightness.}
   \label{V1.5hist}
\end{figure*}

\begin{figure*}[ht!]
   \centering
   \includegraphics[width=1\textwidth]{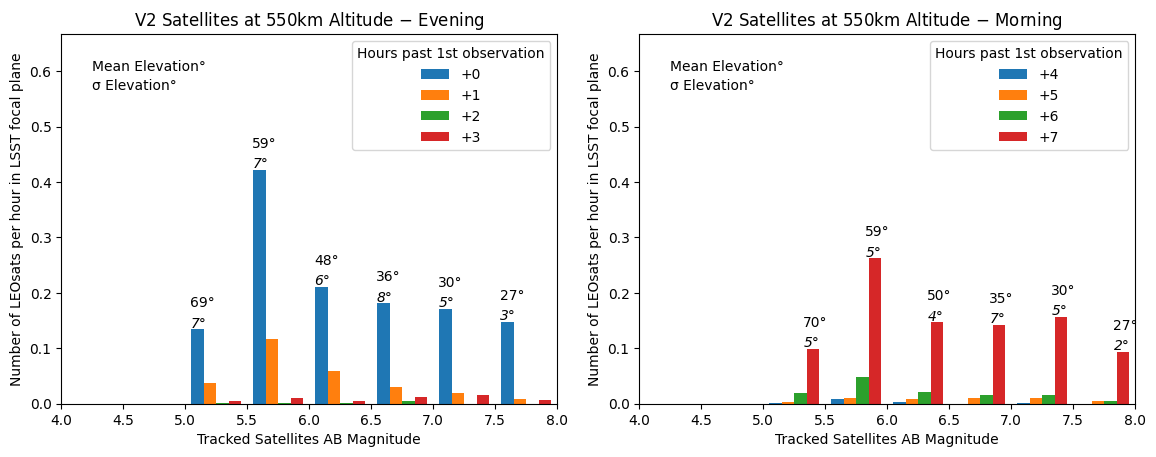}
   \caption{Density (per-thousand on-orbit satellites) of expected bright V2 satellites by the hour since the start of observation. Most bright satellites are seen within the first hour of observation and the final hours of the observation in morning twilight. The mean brightness of all satellites within the first hour is 7.31 AB magnitude with most satellites being between 5.5 - 6 AB magnitude. We predict 0.93 per thousand satellites brighter than 7 AB magnitude in the first hour of observation. And occasionally, a satellite brighter 5.5 AB is encountered. Similar to the previous generation of satellites, the brightest satellites occur at a much higher elevation than the dimmer satellites}
   \label{V2hist}
\end{figure*}

This enables us to compare the two types of satellite and evaluate the improvements made to the newer satellite models. After the simulation yields a database of trails, we categorize it by brightness and the time the trail occurs relative to the first observation of the day. Figure \ref{V1.5hist} and Figure \ref{V2hist} is the visualization of this database for an average day formatted as a histogram for the V1.5 and the V2 Starlink satellites, respectively. This shows the brightness mitigation techniques attempted by SpaceX. These mitigations are described in detail in \citep{spacex_best}.

The number of bright V2 Starlink Walker constellation satellites seen by LSST is significantly lower than the V1.5 Starlink satellites. The mean elevation of the bright V1.5 Starlink satellites is approximately the same as the bright V2 Satellites. However, the variance is smaller for the V2 Starlink satellites, implying that LSST observations of them  are concentrated at higher elevations, while the bright V1.5 Starlink satellites are found in a larger area of sky. Looking at the brightness distribution of V1.5 Starlink satellites in Figure \ref{V1.5hist}, we know that there are two areas of the sky where these satellites are bright. That explains why the bright V1.5 Starlink satellite’s mean elevation is between the two brightness peaks. The V2 Starlink satellite has a fainter peak apparent brightness at low altitudes, so most bright satellites crossing the LSST focal plane are found mostly at higher elevations. 

Satellites are usually illuminated at the beginning and end of the night when the sun altitude is the closest to the horizon. This is a product of two effects: (1) larger volume at orbit height seen in projection close to the horizon, and (2) more specular reflections per satellite at those low AOI.

\begin{figure}[ht!]
   \centering
   \includegraphics[width=0.4\textwidth]{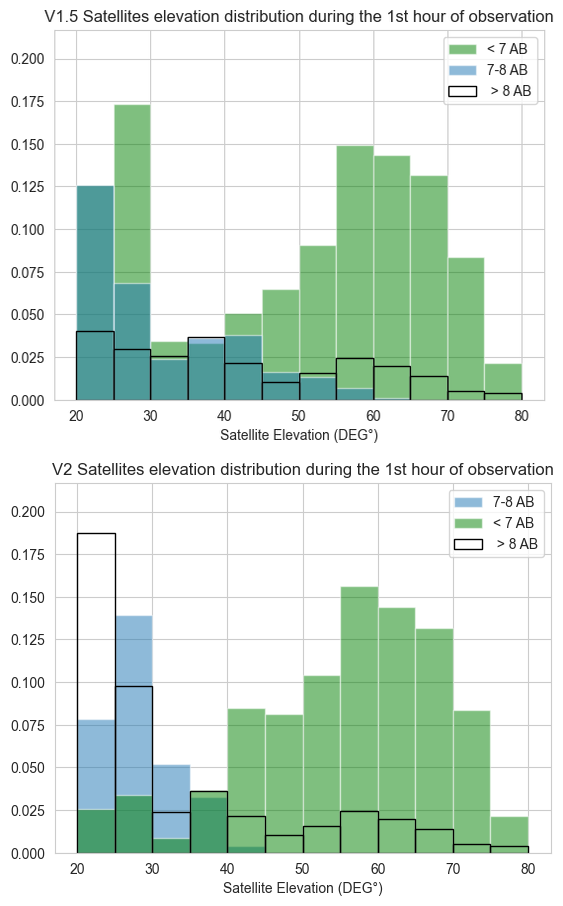}
   \caption{Simulated density per-thousand on-orbit satellites. $Top$: V1.5 Starlink satellites elevation distribution during the first hour of LSST observations on an average summer night. Satellites brighter than 7 AB are found at all satellite elevations, but concentrated at elevations of 20°-30° and 55°-65°. These are the two brightness peaks caused by specular reflection of the chassis and the back scatter from the solar array. On average, there are 0.33 per thousand satellites brighter than 7 AB between 20°-30° elevation. Satellites between 7-8 AB magnitude are observed less frequently, but they are still present at lower satellite elevations. $Bottom$: V2 Starlink satellites elevation distribution during the first hour of LSST observations on an average summer night. 
   Satellites brighter than 7 AB magnitude are observed most frequently high satellite elevations. Unlike the previous versions, the number of satellites brighter than 7 AB are observed at a much lower rate at elevations of 20-30°. In contrast to the 0.33 per thousand V1.5 Starlink satellites brighter than 7 AB between 20°-30° elevation, only 0.083 per thousand V2 satellites are found under the same conditions on an average summer night. This implies that the number of satellites that specularly reflect light is significantly reduced when the V2 Starlink satellites are used in the constellation instead. In fact, most V2 Starlink satellites between 20°-30° elevation that are in position to be seen by the telescope are dimmer than 8 AB magnitude.
   }
   \label{elevationdistribution}
\end{figure}

To further confirm that the number of bright V2 Starlink satellites caused by the chassis’ specular reflection is reduced when compared to the previous generation, we can look at the average elevation distribution of observed satellites. The bright V2 satellites in Figure~\ref{elevationdistribution}  are concentrated at higher satellite elevation while the bright V1.5 satellites are found throughout the sky at all elevations. Since the brightness mitigations by SpaceX reduce the angles at which the sunlight will specularly reflect off the satellite chassis, the LSST scheduled observations do not frequently encounter such satellites.

\bigskip

\begin{figure}[h!]
   \centering
   \includegraphics[width=0.47\textwidth]{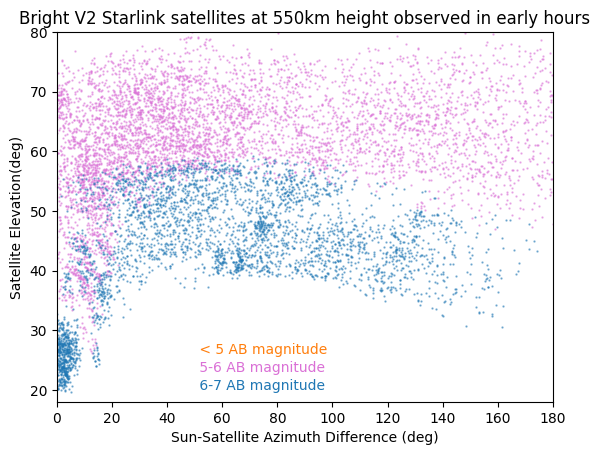}
   \caption{Simulated {V}2 Starlink Satellites observed in the first hours of observation during summer evenings. The model’s dependence on field elevation is apparent. The V2 satellites are split according to their brightness, but a trend emerges: at higher elevations the satellites increase in brightness. Since this plot consists of all sun altitudes in the first two hours of observations, the satellites that are in the earth’s shadow are found at all elevations. However, the fainter satellites are most prominent at low elevations during the beginning of the observation period.}
   \label{v2earlyhours550km}
\end{figure}

When looking at all observations by the LSST during the early hours of the evening in Figure \ref{v2earlyhours550km}, bright {V}2 satellites can only be seen at high elevations and at all sun-sat azimuth differences. In fact, the brightness of the satellite depends on the elevation such that there are no satellites between 6-8 AB magnitude at the higher elevations. Satellites between 6-8 AB magnitude are found at all other elevations making this category of satellites the most common to be observed by the LSST. While most observations are at telescope elevations between 20°-30° and in the direction of the sun, the sun has not set enough for a specular reflection to occur in those places. Therefore, the brightest V2 Starlink satellites can be seen only at high elevations. In fact, the LSST observing program which searches for potentially hazardous asteroids encounters the largest number of bright {V}2 Satellites, but none of those observations are from a specular reflection. That program looks in the direction of the sun and at low telescope elevations, but almost none of those observations are of a {V}2 Starlink brighter than 6 AB magnitude.

\newpage

\subsection{V2 Satellites at 350 Km orbit}
Starlink has begun operating more than 300 satellites below inhabited space stations. These operations are motivated by goals to make space operations safer, improve user service, and mitigate impact on optical astronomy. In this section, we simulate the impact of constellation operations at 350 km altitude on LSST observations.

There are a number of competing effects to consider. Decreased range increases brightness by an $r^2$ law. However, satellites are illuminated for a smaller fraction of their orbit and are visible to fewer observers. Additionally, trail brightness decreases because the satellite moves faster across the sky.

\begin{figure*}[ht!]
   \centering
   \includegraphics[width=1.0\textwidth]{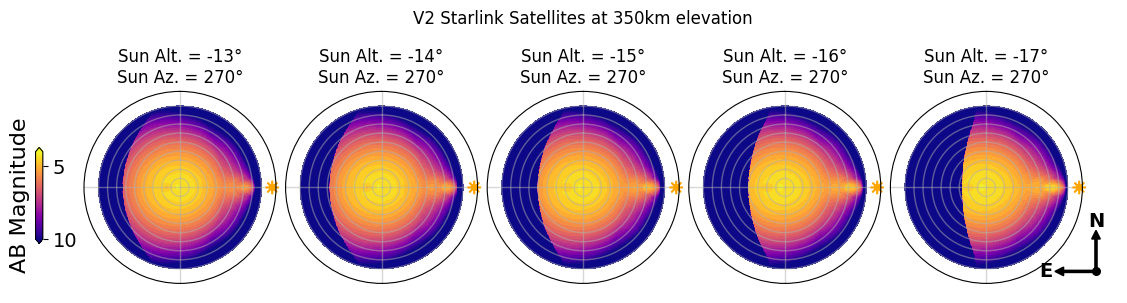}
   \caption{Simulated {V}2 Starlink satellites observed at 350 Km altitude orbit as viewed by the observer on Earth. As expected, the satellites are much brighter than when the satellites were at 550Km. However, the satellites are closer to the surface of Earth, and are expected to go into the Earth's shadow much sooner. While the average tracked brightness of all the satellites visible would be higher, the number of these bright satellites per unit time will be lower.}
   \label{obsview350}
\end{figure*}

Another effect is that satellites in lower orbits encounter the Earth's shadow earlier. This reduces the brightness driven by nadir surfaces. The corresponding sky plot is shown in Figure~\ref{obsview350}.

The effect on the number of satellites as a function of tracked apparent magnitude in the LSST focal plane during scheduled observations is shown in 
Figure~\ref{lowering} in which the histograms for 550km and 350km are placed side by side.  This is a product of satellite brightness and LSST observing schedule.

\subsection{Simulation noise}
Naturally these histograms also have error arising from variations in satellite brightness. Such variance in apparent satellite brightness effectively smears out the magnitude assigned to each histogram bar. While the distribution is not gaussian, we can characterize this simulation error by measuring the full width at half maximum (FWHM) of the satellite apparent magnitudes.

{For the population of satellites that are in the 550km orbital height constellation brighter than 8 AB mag, the FWHM is 0.7 AB mag. Similarly, for the population of satellites that are in the lower orbit at 350km brighter than 8 AB mag, the FWHM is 1 AB mag.} {This ``noise'' in the simulation originates in the real variations in satellite measured brightness. Additional simulation noise is negligible by comparison.} The effect is to scatter the number of satellites predicted towards brighter magnitudes by $\sim$1 magnitude. For example, in Figure~\ref{V2hist} evening histogram this increases the net number of bright satellites by 20\%.  
\subsection{Relative precision and the origin of the faint limit}

It can be seen that there is an advantage to moving lower. This is true for both astronomy and satellite operators. For example, 39\% fewer satellites brighter than 7th mag are seen at the lower 350km orbit. 

The goal of 7th apparent (tracked) magnitude comes from requirement that the electronic ghosts of satellite trails in the LSST CCDs be mitigated via calibration - a process that has a limited dynamic range~\citep{walker+hall2020, tyson2020mitigation}. 7th magnitude is also the limit of unaided eye detection in a dark site. The surface brightness of the faint galaxies in LSST imaging are typically 100 million times fainter than a 7th mag (tracked) satellite. 
\begin{figure*}[ht!]
    \centering
    \includegraphics[width=1.0\textwidth]{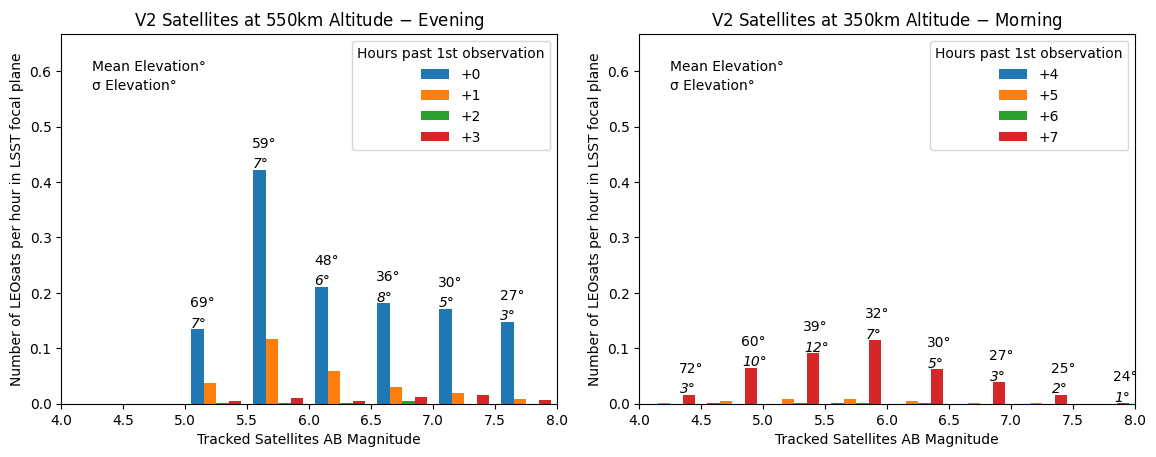}
    \caption{\small The result of moving a simulated Starlink {V2} constellation down from 550km to 350km. The density (per-thousand on-orbit satellites) of satellites in the LSST focal plane per histogram bin per hour are shown vs their tracked magnitude as well as satellite elevation. This is the product of scheduled observations and satellite density. The number (from multiplying the density times the constellation size in units of 1000) in all bins brighter than 7th magnitude must be added to get the total. Variance in satellite brightness imposes some error which is relatively small compared with the overall effect of moving to lower orbits.  It is interesting that brighter Starlinks are forecast as high as 60 deg elevation. These have been reported by observers.  Overall, there is an advantage to moving lower for both astronomy and satellite operators.}
    \label{lowering}
\end{figure*}

\begin{figure}[h!]
   \centering
   \includegraphics[width=0.47\textwidth]{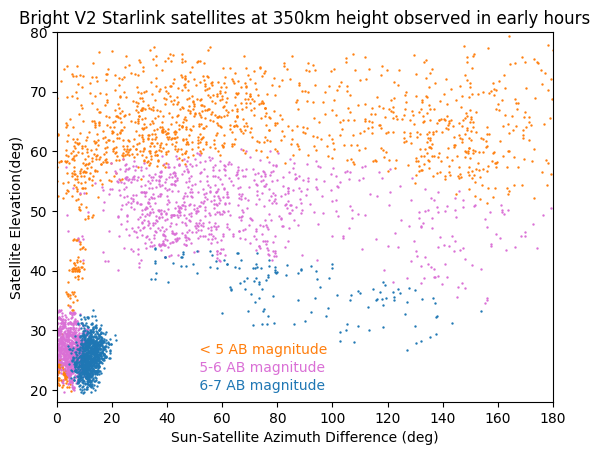}
   \caption{Simulated {V}2 Starlink Satellites at 350km orbits observed in the first hour of observations during summer evenings. Similar to the Starlink satellites at 550km height, there is a dependence on the satellite's elevation. Among the visible satellites, satellites appear brighter at higher elevation. Satellites in the direction of the sun still remain among the brightest. Compared with the 550km simulation, the range to these 350km satellites when overhead is lower so they appear brighter for high elevation.} 
   \label{gen2_350km_first_hour_observations}
\end{figure}

As in Figure~\ref{v2earlyhours550km}, we can plot the corresponding elevation-SSAD-magnitude relation for this lower constellation. We show this in Figure~\ref{gen2_350km_first_hour_observations} for these observed in early hours after twilight. Satellites in the direction of the sun still remain among the brightest. Compared with the 550km simulation, the range is smaller to these 350km constellation satellites when they are observed overhead so that many of them appear brighter.  Since more of these lower satellites encounter the Earth's shadow earlier in the evening, these brighter satellites impact less of the observing program as is apparent in Figure~\ref{lowering}. 
This is not the only observational effect when lowering satellite orbits. There are two additional effects: (1) images of satellite trails in lower orbits spend less time per pixel, and (2) an additional decrease in the number of detected satellites in lower orbits due to a defocus effect.  We review both of these in the discussion section below.


\section{Discussion}

 Tracked (stationary) magnitude and apparent surface brightness each play a role in mitigating interference with optical astronomy. The number of satellites entering some given spot on the sky (such as a camera's field of view) is cast in the language of tracked magnitude. This makes it straightforward to predict when a satellite of known luminosity may present a challenge for observers. The luminosity can be calculated from the integrated BRDF, the range to the satellite, and the sun's illumination of the satellite.  From the number of such satellites in a constellation one can then predict how many satellites will trail across the focal plane of a sky survey per unit time during the night. 
 
 As reported above, when the model constellation is moved to lower orbits, a somewhat lower number of satellites enter the LSST focal plane during observation operations. This is only part of the story. Astronomical detection algorithms are dependent on surface brightness (brightness per pixel). Below some surface brightness threshold, astronomical objects and satellite trails are not efficiently detected. The position-time effects calculated in this work give a decrease of $\sim 40$\% in the number of satellites entering the LSST focal plane.  Of those that do, their peak surface brightness is only 5\% larger.
 
 There are two independent satellite range effects which affect apparent surface brightness. First, as mentioned above, satellites in lower orbits move faster across the camera focal plane, spending less time on each pixel and lowering the trail surface brightness. For satellites observed overhead
they will appear brighter because of the lower range (a $1/r^2$ effect). However, the image of the satellite moves across the LSST focal plane faster (proportional to $r$). The net effect is that the surface brightness of the trail increases only as $1/r$.  
This offsets the trend to brighter {\textit{tracked magnitudes}} reported above in Figure~\ref{gen2_350km_first_hour_observations}.
Note that while there are some brighter tracked satellites at 350 km compared with 550 km, the satellites at 350 km move faster across the focal plane thus generating lower apparent surface brightness.
 
 Second, as LEO satellite orbits are lowered, they become more out of focus in cameras with large aperture telescopes. For example, in Rubin Observatory's LSST moving satellites from 550km to 350km lowers the surface brightness. Compared with an in-focus star, there is a factor of 6 decrease in trail surface brightness for satellites at 350km~\citep{tyson2024expected}.  
 
 All these effects contribute to the apparent surface brightness of a satellite with fixed BRDF as its orbit is changed. The faster apparent angular speed and more blurred image at lower orbits largely cancels the increase in brightness due to the smaller range.  As an example, for LSST observing a satellite overhead, the peak surface brightness of a model 7 $\times$ 3m satellite at 350km is only 5\% brighter that at 550km. These defocus effects on the observed satellite trail optical surface brightness are discussed in greater detail in~\citep{snyder2025satellite}.

 Taken together, the decreased number of 350km simulated satellites and their nearly equal surface brightness compared with the simulated 550km constellation, indicate a clear advantage for the lower orbit.

 \subsection{Impact on LSST science}

The impact of such trails on the survey science then may be understood by converting the tracked magnitude of the satellite to the calibrated apparent surface brightness, and then examining the impact of that surface brightness on the camera and corresponding science program. Sufficiently faint trail surface brightness will not impact the science.  The threshold for this is a function of the science program.   Calibrated lab measurements on the sensors used in the survey has informed the numerical value of this critical surface brightness. 
 
 The science impact also scales with the number of visible satellites. This impact falls into two broad categories: the search for rare transient objects in the universe (the ``time domain''), and the ultra deep sky science (cosmology, galaxy research). As discussed in~\citep{fankhauser2023satellite}, momentary reflections of the sun off satellite hardware may cause optical flares or glints which can masquerade as real astronomical transient objects. In this study, these estimates have been made via models of satellite brightness together with LSST observing cadence. As shown above in Figure~\ref{lowering}, the number of these visible satellites decreases as a constellation is moved to a lower orbit. 

 The overall science impact on LSST of course depends on the whole satellite population and not just Starlink. It is the hope of the authors that other operators also adopt effective darkening mitigations across the electromagnetic spectrum, and this Starlink-focused study demonstrates that off-pointed solar arrays, reduced diffuse reflections, and lower orbital altitudes can be effective ways to achieve this for large optical surveys like LSST.

 \bigskip

\newpage
\section*{Acknowledgments}
We acknowledge support from NSF/AURA/LSST grant N56981CC, NSF grant AST-2205095, and DOE grant DE-SC0009999 to UC Davis. We acknowledge useful discussions with Meredith Rawls, Adam Snyder, Craig Lage, Lynne Jones, Daniel Polin, Cole Morgan, David Goldstein, Andrew Connolly, and Jeff Shaddix. We acknowledge Slingshot Aerospace for providing contracted brightness measurements of Starlink satellites on-orbit. 
Software: Astropy \citep{astropy2022}, NumPy \citep{numpy2020}, SciPy \citep{scipy2020}, Matplotlib \citep{matplotlib2017}, pandas \citep{pandas2010}, SGP4, Celestrak, Space-Track.

\bigskip
\section*{Data Availability}
The software written for this project as well as the data are available at two open Github repositories: \url{https://github.com/Phan-Kandula/simulated-satellite-streaks-impact/}
and a permanent archive can be found on Zenodo \citep{datapipeline2025}. \texttt{Lumos-Sat} is an open-source Python package that was created for this project \citep{lumos2023}. Documentation and installation instructions for \texttt{Lumos-Sat} can be found at: \url{https://lumos-sat.readthedocs.io/}. Development of \texttt{Lumos-Sat} is ongoing and collaboration is encouraged.
\bibliography{main}{}
\bibliographystyle{aasjournal}
\end{document}